\documentclass[5p]{elsarticle}
\usepackage[english]{babel}
\usepackage[applemac]{inputenc}
\usepackage{bbm}
\usepackage{amsthm,amssymb,amsmath}
\usepackage[toc]{appendix}
\usepackage{color}

\newcommand{\qsp}[2]{\,\ensuremath{\raise.5ex\hbox{$#1$}\big\slash\raise-.5ex\hbox{$#2$}}}

\newcommand{\Tr}[1]{\mathrm{Tr}\left[#1\right]}

\newcommand{\dl}{\mathsf{d}_\ell} 

\renewcommand{\P}{\mathcal{P}}

\newcommand{\id}{\mathbbm{1}}
\newcommand{\Ad}{\mathsf{Ad}}

\newcommand{\ad}{\mathsf{ad}}

\begin{document}

\title{{Symmetric logarithmic derivative for general $n$-level systems and the quantum Fisher information tensor for three-level systems}}
\author[erc]{E. Ercolessi\corref{cor2}}
\ead{ercolessi@bo.infn.it}
\author[sch]{M. Schiavina\corref{cor1}\fnref{telMS}}
\ead{michele.schiavina@math.uzh.ch}

\cortext[cor2]{Corresponding author} 
\cortext[cor1]{Principal corresponding author}
\fntext[telMS]{Telephone: $+41(0)446355872$, Fax: $+41(0)446355705$}

\address[erc]{Dipartimento di Fisica e Astronomia, Universit\`a di Bologna and INFN, Sezione di Bologna, via Irnerio 46, 40126 Bologna, Italy}
\address[sch]{Institut f\"ur Mathematik, Universit\"at Z\"urich, Winterthurerstrasse 190, 8057 Z\"urich, Switzerland}

\begin{abstract}{ Within a geometrical context, we derive} an explicit formula for the computation of the symmetric logarithmic derivative for arbitrarily mixed quantum systems, provided that the structure constants of the associated unitary Lie algebra are known. \\
To give examples of this procedure, we first recover the known formulae for two-level mixed and three-level pure state systems and then apply it to the novel case of $U(3)$, that is for arbitrarily mixed three-level systems (q-trits).\\
Exploiting the latter result, we finally calculate an expression for the Fisher tensor { for a q-trit considering also all possible degenerate subcases}.

\end{abstract}

\maketitle

\section{Introduction}\label{se:quantum}

The understanding of the geometrical structures underlying quantum mechanics \cite{erc} allows to get precious insights in many physical phenomena, ranging from the well-known Berry phase \cite{berry,aa} and \cite{rez} to recent developments in quantum information theory \cite{bro} and entanglement problems \cite{beng}.
A recent  instance of this is given by the geometric interpretation of the so-called quantum Fisher information index. It can be shown \cite{brau,brauII,hel,hol} to be an upper bound for the amount of information one can extract by making measurements on a quantum state, setting thus a problem of optimization \cite{brau,brauII,barn,luati,es}.

If a family of (pure or mixed) quantum states is given through a density matrix $\rho(\theta)$ depending on a a single parameter $\theta$, one can define the quantum Fisher information as:
\begin{equation}
I(\theta)=\Tr{\rho(\dl\rho)^2}
\end{equation}
where the so-called symmetric logarithmic derivative $\dl\rho$ appears. The latter is implicitly defined through the formula:
\begin{equation}\label{symlog}
d\rho=\frac{1}{2}\left\{\rho,\dl\rho\right\}
\end{equation}
with $d\equiv d_\theta$ denoting the derivative with respect to $\theta$.

It was first Rao, as back as in 1945, who first made a connection between Fisher information and the geometry of the parameter-space submanifold in the the Hilbert space \cite{rao}. { Let us recall that the Fisher information is not the only metric one can put on the space of density matrices, which can be endowed with many distances, useful in several situations \cite{petz,zyc}.} Since Rao, many developments have been made to understand the nature of the Fisher index, by looking at the problem in many different contexts \cite{bur,bro2,bro3,brohug}. 
A geometric interpretation kicks in as soon as one recognizes that the space of states for finite dimensional quantum systems can be seen as the collection of \emph{co-adjoint orbits of the unitary group} with unitary trace \cite{chat,es}. Each of these orbits being a submanifold of the unitary Lie algebra,  corresponding to a particular choice of the mixing parameters. In virtue of this identification, much effort has been spent in showing that the Fisher information index is closely related to the metric that one can endow these manifolds with. In particular in \cite{brohug,marmo} it is shown that for the $n$-dimensional pure state case the Fisher information index, seen as a metric, is exactly the Fubini-Study metric on the complex projective space of $n$-dimensional, rank-one, idempotent, self adjoint density matrices. This result was extended to two-dimensional mixed states systems in \cite{es}, where it is shown that the Fisher information metric is equivalent to the round metric on $\mathcal{S}^3$, that is the complete space of states for twofold mixed $2$-level quantum systems (q-bits), and it yields a weighted version of the Fubini-Study metric when the mixing parameters are kept fixed. { Another interesting geometric interpretation of these objects can be found in \cite{Fuji}, where quantum channels (dynamical evolutions of quantum systems) are rewritten in the language of fibre bundles.}

It is evident that a nontrivial constituent in the definition of the quantum Fisher index is the symmetric logarithmic derivative, which is only implicitly defined through Eq. \eqref{symlog}. The aim of this Letter is to present a general method to transform the problem of calculating the symmetric logarithmic derivative into that of solving a set of linear algebraic equations. {This result is of particular interest because the issue of computing the Fisher information index for arbitrary $n$-dimensional states becomes hard for large $n$. The original methods proposed by Braunstein and Caves \cite{brau} become in this case non-optimal, for they rely on the eigenstate decomposition of the main quantities $d\rho$ and $\dl\rho$.

Moreover, up to now the only computable higher dimensional case was the pure state case, but it is a fact that decoherence processes in realistic evolving physical models force the outcoming states to be mixed. A method to compute relevant quantities such as the Fisher information index that overcomes the original difficulties is then surely of some interest to the community.}

 As we will show in sect. \ref{se:slf}, this method rests just on the algebraic properties of the Lie algebra of $U(n)$ and may in principle be applied for any $n$-level system, both for the pure and the mixed cases. { As a matter of fact it is sufficient to give a parameterization of some space of states $O$ and exploit the proper tensorial nature of the key objects: $d\rho,\dl\rho\in\Omega^1(O)\otimes \mathfrak{u}(n)$. General solutions to the parametrization problem of this kind of manifolds is presented e.g. in \cite{pick}, while some higher dimensional $\mathfrak{su}(n)$ generators and structure constants may be found in \cite{sarid,Maki}. Notice furthermore that the problem of finding the structure constants for $U(n)$ and a representation of its Lie algebra is interesting \emph{per se}. Therefore, the required elements for this method to work are not \emph{ad hoc}, and do not suffer of theoretical obstructions in their computation.}
 
 To show how this method works, we calculate explicitly the symmetric logarithmic derivative for a $3$-level system, for the pure and all mixed subcases. Finally, in sect. \ref{se:ft} we exploit this result to explicitly calculate the so-called Fisher tensor for q-trits and establish a connection with the results found in \cite{es} for a $2$-level system.

\section{Symmetric logarithmic one form} \label{se:slf}
Since the quantum Fisher information is dependent on the symmetric logarithmic derivative, it is of capital importance to have an explicit formula for this implicitly defined object. We would like to present here a formula that solves the problem in general, anytime the structure constants of the underlying Lie group are known, reducing the problem of finding the correct expression for $\dl\rho$ to a simple problem of linear algebra. In this section we will drop the dependence of our object on $\theta$, moreover, we will hereinafter consider $d$ as the  differential acting on matrices in the Lie algebra like $\rho$. Thus, the symmetric logarithmic derivative $\dl\rho$ together with the ordinary derivative $d\rho$ will be regarded as sections of the cotangent bundle, that is to say one forms on the space of states, with values in the Lie algebra.

\subsection{Lie algebra expansion}\label{formula}
To any given rank-m mixed quantum state of an $n$-level system, represented by the $n\times n$ density matrix
\begin{equation}
\rho=\sum_{i=1}^{m}k_i P_i,
\end{equation}
where $P_i^2=P_i=P_i^\dag$ and $\sum_i k_i =1,\ k_i\geq0$, it is possible to associate \cite{es,beng,zyc} the orbit under the (co-)adjoint\footnote{Here we are exploiting the canonical identification of the Lie algebra and its dual, justified by the fact that we are working with a compact group of matrices.} action of the unitary group
\begin{equation}
\P^{(m)}_n=\{\rho\in\mathfrak{u}(n)\ |\ \rho=\Ad_U\rho_0=U^\dagger \rho_0 U,\ U\in\mathsf{U}(n)\},
\end{equation}
passing through the point
\begin{equation}
\rho_0=\mathsf{diag}_n\{k_1\dots k_m, 0\dots 0\}.
\end{equation}
This implies that it is possible to describe any point $\rho$ of the orbit by means of an expansion in the Lie algebra generators of $\mathfrak{su}(n)$ plus the identity matrix. Moreover, the base point matrix $\rho_0$ is diagonal and has an expansion limited to the diagonal generators in the Lie algebra, which are $n-1$ and will be denoted by hatted indices. Therefore we can set:
\begin{equation}\label{basepointexpansion}
\rho_0=\rho_\id\id + \sum_{\widehat{l}=1}^{n-1}\rho_{\widehat{l}}\, t_{\widehat{l}}=\rho_\id\id + \sum_{k=1}^{n^2-1}\rho_{k} t_{k},
\end{equation}
where the $t_i$'s are the Lie algebra generators in the $n$-dimensional fundamental representation\footnote{Notice that diagonal generators $t_\mu$, labeled with hatted indices, may be found also in the set of all generators, labeled with roman indices. As a matter of fact, $t_{\mu=1}$ is usually found as $t_{i=3}$. }, normalized in such a way that $\Tr{t_it_j}=2\delta_{ij}$.
A similar expansion can be provided also for the one forms:
 \begin{equation}\label{basepointexpansion2}
\begin{aligned}
d\rho_0=&D_\id \id + \sum_{i=1}^{n^2-1}D_i t_i,\\
\dl\rho_0=&L_\id \id +\sum_{i=1}^{n^2-1}L_i t_i. 
\end{aligned}
\end{equation}
the differential part being borne by the one forms $D_i$ and $L_i$. { Notice that the parameters $k_i$ are not allowed to vary, for the time being. When we will consider relaxing this constraint, it will be appropriately specified.} 
Expressions of the matrices $\rho,d\rho,\dl\rho$ above other points in the orbit may be found by adjoint-acting on the base point expressions given above. This is true in virtue of the fact that the inclusion map
\begin{equation}
\iota\colon \P^{(m)}_n\longrightarrow \mathfrak{u}(n)
\end{equation}
sending a point $\rho$ of the submanifold in the Lie algebra is naturally equivariant with respect to the adjoint action of $U(n)$. Therefore its differential
\begin{equation}
d\iota\colon T\P^{(m)}_n\longrightarrow \mathfrak{u}(n)
\end{equation}
is again equivariant. Thus, the differential yields an equivariant Lie algebra valued one form $d\rho=\Ad_Ud\rho_0$, and this is sufficient to show that $\dl\rho$ must be equivariant as well.

An explicit way to view this at the level of tangent vectors is considering curves in $\mathfrak{u}(n)$ passing through $\rho_0$. { Namely we have that $\rho_0(\theta)=\sum k_iP_i(\theta)$ and the derivative w.r.t. $\theta$ reads 
\begin{equation}
\partial_\theta\rho_0(\theta)= \sum k_i\partial_{\theta}  P_i(\theta)
\end{equation}
Then, one can act with the co-adjoint action (independent of $\theta$) and find a tangent vector above any other point, namely:
\begin{equation}
X_\rho=\partial_\theta U\rho_0(\theta)U^\dag = U\partial_\theta\rho_0(\theta)U^\dag=UX_{\rho_0}U^\dag
\end{equation}
where $X_\rho$ and $X_{\rho_0}$ represent the tangent vectors to the orbit over $\rho$ and $\rho_0$ respectively, and are in one to one correspondence with the differentials $d\rho_0,d\rho$. This argument works also for $\theta$-dependent parameters $k_i(\theta)$: indeed in this case one has
\begin{equation}
\partial_\theta\rho_0(\theta)= \sum \partial_\theta k_i(\theta) P_i(\theta) +\sum k_i(\theta) \partial_\theta P_i(\theta)
\end{equation}
where the first term, which contains only derivatives of the scalar coefficients, commutes with $\rho_0(\theta)$ for each value of the parameter.  This means that, with respect to the algebra decomposition introduced above, it involves only the diagonal generators. We will come back to this point in section \ref{transv}.}

Going back to the point $\rho_0$, plugging the \eqref{basepointexpansion2}'s into \eqref{symlog} one finds
\begin{equation}\begin{aligned}
D_\id \id + \sum_{l=1}^{n^2-1}D_l t_l=\left(L_\id\rho_\id+\frac{2}{n}\sum_{j=1}^{n^2-1}L_j\rho_j\right)\id\\ 
+ \sum_{l=1}^{n^2-1}\left( \rho_\id L_l+L_\id\rho_l + \sum_{j,k=1}^{n^2-1}L_j\rho_kf_{jkl}\right)t_l
\end{aligned}\end{equation}
where the $f$'s symbols are defined through the relations: $\{t_j,t_k\}=\frac{4}{n} \delta_{jk}\id + 2\sum_kf_{jkl}t_l$. This yields the following set of equations, relating the coefficients in the expansions  of the ordinary and logarithmic derivatives of $\rho_0$:
\begin{equation}\label{defeq}\begin{aligned}
D_\id&=\rho_\id L_\id+\frac{2}{n}\sum_{j=1}^{n^2-1}\rho_jL_j \\
D_l&=\rho_\id L_l+\rho_l L_\id + \sum_{j,k=1}^{n^2-1}\rho_kL_jf_{kjl},
\end{aligned}
\end{equation}
As a general feature of the co-adjoint orbits picture, it is possible \cite{kir,es} to express the tangent (co-)vector {$d\rho_0$ as a commutator of $\rho_0$} and a Hermitian matrix\footnote{Properly a (co-)vector, or a one form, with values in the Lie algebra.} $K_0=K_\id\id+\sum_{l=1}^{n^2-1}K_lt_l$:
\begin{equation}
d\rho_0=\left[K_0,\rho_0\right] . 
\end{equation}
Defining now $[t_i,t_j]=2i\sum_kc_{ijk}t_k$ (with diagonal generators commuting with each other), we may write:
\begin{equation}
d\rho_0=2i\sum_{k,i,\bar{l}=1}^{n^2-1}c_{ki\bar{l}}K_k\rho_i t_{\bar{l}}, 
\end{equation}
where the barred indices run over non-diagonal generators. Therefore we find:
\begin{equation}\begin{aligned}\label{indeterminacy}
D_\id&=\rho_\id L_\id+\frac{2}{n}\sum_{j=1}^{n^2-1}\rho_jL_j=0\\
D_{\widehat{l}}&=\rho_{\widehat{l}}L_\id+ \rho_\id L_{\widehat{l}} + \sum_{j,k=1}^{n^2-1}\rho_kL_jf_{kj\widehat{l}}=0.
\end{aligned}\end{equation}
To find the explicit expression for the symmetric logarithmic form { $\dl\rho_0$} it is then sufficient {to solve the following $n^2$ equations for the $n^2$ unknown $L_j$'s}:
\begin{equation}\label{defining}
D_l=\rho_\id L_l+\rho_l L_\id + \frac{1}{2}\sum_{j,k=1}^{n^2-1}\rho_kL_jf_{kjl},
\end{equation}
while the set of solutions to \eqref{indeterminacy} yields free choices on the definition of $\dl\rho$, which is indeed manifestly defined in \eqref{symlog} up to matrices that anticommute with $\rho$.

Notice however that when $\rho_0$ is full rank and positive, that is to say $k_i>0$, there is no Hermitian matrix anti-commuting with $\rho_0$. This means that the solution to the affine problem \eqref{symlog} has a unique solution $\dl\rho$. Moreover, looking at the homogeneous equations \eqref{indeterminacy} it is easy to gather that these are equations for the diagonal components $L_{\widehat{i}}$ of $\dl\rho_0$. As a matter of fact we have that $\rho_jL_j=0$ if $j$ refers to a non-diagonal generator because $\rho_0$ is diagonal, moreover, the term $\rho_{\widehat{k}}L_jf_{\widehat{k}j\widehat{l}}=0$ whenever $L_j$ is a non-diagonal component because
\begin{equation}\label{structvanish}
f_{\widehat{k}j\widehat{l}}\propto\Tr{\{t_{\widehat{k}},t_j\}t_{\widehat{l}}}=0
\end{equation}
So the unique solution to \eqref{indeterminacy} is the trivial one and hence there are no diagonal components for $\dl\rho$.

As a check of our formulae, we show how to recover the expressions for the symmetric logarithmic form in the case of 2 dimensional pure \cite{barn,marmo} and mixed states \cite{luati,es}. In these cases, the density matrix is a $2\times 2$ Hermitian matrix and $\rho_0=\mathsf{diag}(k_1,k_2)$, with $k_1+k_2=1$. The generators of the Lie algebra are given by Pauli matrices $\sigma_{1,2,3}$ and the identity matrix. Therefore we have the expansion:
\begin{equation}
\rho_0=\frac{k_1+k_2}{2}\id + \frac{k_1-k_2}{2}\sigma_3
\end{equation}
It is easy to solve Eqs. \eqref{indeterminacy} in this case, because all symmetric structure constants $f_{ijk}$ vanish and we are left only with
\begin{equation}\begin{aligned}
D_3&=\frac{1}{2}L_3+\frac{k_1-k_2}{2}L_\id=0\\
D_\id&=\frac{1}{2}L_\id+\frac{k_1-k_2}{2}L_3=0
\end{aligned}\label{indeterminacy2}\end{equation}
For mixed states ($k_2\not=0$) this yields $L_\id=L_3=0$ meaning that the symmetric logarithmic form is uniquely defined as discussed above. Then, equations \eqref{defining} reduce simply to
\begin{equation}
D_l=\rho_\id L_l=\frac{k_1+k_2}{2}L_l
\end{equation}
which agree with what was found in \cite{es,luati,barn,marmo},  yielding
\begin{equation}\label{twolevel}
\dl\rho=\frac{2}{k_1+k_2}\left(D_1t_1 + D_2t_2\right)
\end{equation}
It is worthwhile to notice that the matrix associated to the above homogeneous equations has determinant
\begin{equation}
\mathrm{det}\left(\begin{array}{cc}
\rho_\id & \rho_3 \\
\rho_3 & \rho_\id
\end{array}\right)=k_1k_2
\end{equation}
and that in the pure state case ($k_1=1,k_2=0$) Eqs. \eqref{indeterminacy2} have a one-dimensional set of nontrivial solutions. In this particular case, for any choice of $L_\id$ one has $L_3=-L_\id$, so that the symmetric logarithmic one form is defined uniquely up to diagonal terms of the form
\begin{equation}
\widetilde{D}=
\mathsf{diag}\{0, 2L_\id\}.
\end{equation}
manifestly anticommuting with $\rho_0=\mathrm{diag}\{1,0\}$.

\subsection{Mixed states of a three-level system}\label{3mixed}

To prove the power of our approach for the computation of the symmetric logarithmic form $\dl\rho$, we shall compute the explicit case of a mixed state in $U(3)$, that is of a $3$-level system. This case is described by matrices conjugated via the adjoint action to
\begin{equation}
\rho_0=\mathsf{diag}\{k_1,k_2,k_3\}.
\end{equation}
with as usual $k_1+k_2+k_3=1$ and distinct. The basis we will use to represent the Lie algebra is given by the Gell-Mann matrices \cite{emm}
\begin{equation}\begin{aligned}
\lambda_1=&\left(\begin{array}{ccc}
			 0 & 1 & 0\\
			 1 & 0 & 0\\
			 0 & 0 & 0\end{array}\right),
&\lambda_2=\left(\begin{array}{ccc}
			 0 & -i & 0\\
			 i & 0 & 0\\
			 0 & 0 & 0\end{array}\right)\\
\lambda_4=&\left(\begin{array}{ccc}
			 0 & 0 & 1\\
			 0 & 0 & 0\\
			 1 & 0 & 0\end{array}\right),
&\lambda_5=\left(\begin{array}{ccc}
			 0 & 0 & -i\\
			 0 & 0 & 0\\
			 i & 0 & 0\end{array}\right)\\
\lambda_6=&\left(\begin{array}{ccc}
			 0 & 0 & 0\\
			 0 & 0 & 1\\
			 0 & 1 & 0\end{array}\right),\ \ 
&\lambda_7=\left(\begin{array}{ccc}
			 0 & 0 & 0\\
			 0 & 0 & -i\\
			 0 & i & 0\end{array}\right)\\
\lambda_3=&\left(\begin{array}{ccc}
			 1 & 0 & 0\\
			 0 & -1 & 0\\
			 0 & 0 & 0\end{array}\right),\ \ 
&\lambda_8=\frac{1}{\sqrt{3}}\left(\begin{array}{ccc}
			 1 & 0 & 0\\
			 0 & 1 & 0\\
			 0 & 0 & -2\end{array}\right)
\end{aligned}\end{equation}
satisfying the following relations:
\begin{equation}\begin{aligned}
\{\lambda_j,\lambda_k\}=&\frac{4}{3}\delta_{jk}\id+2\sum_{l=1}^8 f_{jkl}\lambda_l\\
\left[\lambda_j,\lambda_k\right]=&2i\sum_{l=1}^8c_{jkl}\lambda_l
\end{aligned}\end{equation}
where the totally antisymmetric structure constants are:
\begin{equation}\begin{aligned}
c_{123}=1&;\ c_{458}=c_{678}=\frac{\sqrt{3}}{2}\\
c_{147}=c_{246}=c_{257}&=c_{345}=-c_{156}=-c_{367}=\frac{1}{2}
\end{aligned}\end{equation}
while the totally symmetric symbols are given by:
\begin{equation}\begin{aligned}
f_{118}&=f_{228}=f_{338}=-f_{888}=\frac{1}{\sqrt{3}}\\
f_{448}&=f_{558}=f_{668}=f_{778}=-\frac{1}{2\sqrt{3}}\\
f_{146}&=f_{157}=-f_{247}=f_{256}=\frac{1}{2}\\
f_{344}&=f_{355}=-f_{366}=-f_{377}=\frac{1}{2}
\end{aligned}\end{equation}

It is easy to see that $\rho_0$ can be expanded as
\begin{equation}\label{rhozeroexpansion}
\rho_0=\frac{1}{3}\id + \frac{k_1-k_2}{2}\lambda_3 + \frac{k_1+k_2-2k_3}{2\sqrt{3}}\lambda_8
\end{equation}
and that Eqs. \eqref{defeq} become (dropping the summation symbol, understanding sum over repeated indices):
\begin{equation}\begin{aligned}
d\rho_0&=\left(\rho_\id L_\id + \frac{2}{3}(\rho_3L_3 + \rho_8L_8) \right)\id\\
&+ \left(\rho_3L_\id + \rho_\id L_3 + \rho_3 L_k f_{k33}\right)\lambda_3\\
&+\left(\rho_8L_\id + \rho_\id L_8 + \rho_8L_kf_{k88} \right)\lambda_8 \\
&+ \left(\rho_\id L_{\bar{i}} + \rho_3L_kf_{k3\bar{i}}+\rho_8L_jf_{j8\bar{i}}\right)\lambda_{\bar{i}}
\end{aligned}\end{equation}
where again the barred indices run over non-diagonal generators. We can now extract the defining equations for $\dl\rho$ to be:

\begin{equation}\label{equazioniU3}\begin{aligned}
&D_1=\left(\rho_\id + \frac{1}{\sqrt{3}}\rho_8\right)L_1 = \frac{(k_1+k_2)}{2}L_1\\
&D_2=\left(\rho_\id + \frac{1}{\sqrt{3}}\rho_3\right)L_2 = \frac{(k_1+k_2)}{2}L_2\\
&D_4=\left(\rho_\id + \frac{1}{2}\rho_3-\frac{1}{2\sqrt{3}}\rho_8\right)L_4=\frac{(k_1+k_3)}{2}L_4\\
&D_5=\left(\rho_\id + \frac{1}{2}\rho_3-\frac{1}{2\sqrt{3}}\rho_8\right)L_5=\frac{(k_1+k_3)}{2}L_5\\
&D_6=\left(\rho_\id - \frac{1}{2}\rho_3-\frac{1}{2\sqrt{3}}\rho_8\right)L_6=\frac{(k_2+k_3)}{2}L_6\\
&D_7=\left(\rho_\id - \frac{1}{2}\rho_3-\frac{1}{2\sqrt{3}}\rho_8\right)L_7=\frac{(k_2+k_3)}{2}L_7
\end{aligned}
\end{equation}

plus three homogeneous equations:

\begin{equation}
\begin{aligned}
&D_\id=\rho_\id L_\id + \frac{2}{3}(\rho_3L_3 + \rho_8L_8) = 0\\
&D_3=\rho_3L_\id + \rho_\id L_3 + \frac{1}{\sqrt{3}}(\rho_8 L_3 + \rho_3 L_8) = 0\\
&D_8=\rho_8L_\id + \rho_\id L_8 + \frac{1}{\sqrt{3}}(\rho_3 L_3 - \rho_8L_8) = 0 
\end{aligned}
\end{equation}

{\it  {Full rank case.}}\\
As already discussed for the general case, for nonzero values of the $k_i$, the first three homogeneous equations admit the unique trivial solution $L_\id=L_3=L_8=0$. This can be justified also because  the determinant of the associated matrix is
\begin{equation}\label{U3degmatrix}
\mathrm{det}\left(\begin{array}{ccc}
\rho_\id & \frac{2}{3}\rho_3 & \frac{2}{3}\rho_8\\
\rho_3 & \rho_\id + f_{383}\rho_8 & f_{833}\rho_3 \\
\rho_8 & f_{338}\rho_3 & \rho_\id + f_{888}\rho_8
\end{array}\right)=k_1k_2k_3,
\end{equation}
vanishing only when $\rho$ is not full rank.\\
Therefore, the symmetric logarithmic form is easily computed from the Lie algebra expansion of the ordinary differential ${d\rho_0}$ as
\begin{equation} \label{threelevel} \begin{aligned}
{\dl\rho_0}&=\frac{2}{k_1+k_2}(D_1\lambda_1+D_2\lambda_2)
+\frac{2}{k_1+k_3}(D_4\lambda_4+D_5\lambda_5) \\ &+ \frac{2}{k_2+k_3}(D_6\lambda_6+D_7\lambda_7) \end{aligned} 
\end{equation}
If we compare this formula with the one found in \eqref{twolevel} for a two-level mixed state system, it is clear that Eq. \eqref{threelevel} highlights the splitting into three $SU(2)$ components that underlies the group structure of $U(3)$. \\

{\it {Rank 2 mixed non-degenerate case.}}\\
From the above computation it is easy to recover the expressions of $\dl\rho$ for a three-level system that mixes only two of the available states, when we may assume $k_3=0$ and $k_1,k_2\neq 0$ with $k_1+k_2=1$. From formula \eqref{U3degmatrix} it is clear that now we encounter some ambiguities in the definition of $\dl\rho$. In particular, it is just straightforward  to solve the linear system associated to \eqref{U3degmatrix} and obtain that $\dl\rho$ is defined up to matrices of the form
\begin{equation} \label{dtilde3}
\widetilde{D}=L_\id \id - \sqrt{3}L_\id \lambda_8=\mathsf{diag}\{0, 0, 3L_\id \}.
\end{equation}

We can now solve the remaining six equations in \eqref{equazioniU3}:
\begin{equation}
\begin{aligned}
L_{1,2}&=\frac{2}{k_1+k_2}D_{1,2}\\
L_{4,5}&=\frac{2}{k_1}D_{4,5}\\
L_{6,7}&=\frac{2}{k_2}D_{6,7}
\end{aligned}
\end{equation}
yielding the following expression for the symmetric logarithmic derivative:
\begin{equation}
\dl\rho=\dl\rho_{SU(2)}+\dl\rho_{\mathrm{res}} + \widetilde{D}
\end{equation}
where the last term is given in (\ref{dtilde3}), the first one equals the one we have obtained for the $SU(2)$ case in eq. (\ref{twolevel}) { with the generators $t$ replaced by the appropriate $\lambda$'s},
while the middle one has the following form:
\begin{equation}
\dl\rho_{\mathrm{res}}=\left(\begin{array}{ccc}
0&0&\frac{1}{k_1}D^-_2\\
0&0&\frac{1}{k_2}D^-_3\\
\frac{1}{k_1}D^+_2& \frac{1}{k_2}D^+_3 & 0
\end{array}\right)
\end{equation}
with $D_2^{\pm}=D_4\pm iD_5$ and $D_3^{\pm}=D_6\pm iD_7$. \\

{\it {Pure state case.}}\\
From the equations found in the generic 3 level case it is straightforward to obtain the expressions for $\dl\rho$ in the pure state case. One must only pay attention to the degenerations that arise from fixing, for example, $k_1=1,\ k_2=k_3=0$. As a matter of fact, Eqs. \eqref{equazioniU3} become
\begin{equation}\label{dlpure3}
D_i=
\begin{cases}
0 & i=\id,3,6,7,8\\
\frac{1}{2}L_i & i=1,2,4,5
\end{cases}
\end{equation}
In particular it is important to check that $D_6, D_7$ are zero by other means, so to be compatible with the above equations for any value of $L_6, L_7$. 
This follows easily imposing that $d\rho$ be written as a commutator with a matrix $K_0$.

From Eqs. \eqref{dlpure3} it is clear that one possible choice for $\dl\rho$ is 
\begin{equation}
\dl\rho=2d\rho
\end{equation}
yet one remains with a degeneracy in the solution of Eq. \eqref{symlog} tantamount to the freedom of choice of $L_6$ and $L_7$.\\

{\it {Eigenvalue degeneration in q-trits.}}\\
Here we would like to consider the case in which two out of three eigenvalues coincide, by imposing, say, $k_2=k_3$, i.e. $k_1+2k_2=1$. Now we have:
\begin{equation}
\begin{aligned}
D_{1,2}&=\frac{(k_1+k_2)}{2}L_{1,2}\\
D_{4,5}&=\frac{(k_1+k_2)}{2}L_{4,5}\\
D_{6,7}&=k_2L_{6,7}
\end{aligned}
\end{equation}

As it will be shown in detail in the next section, where an explicit complex parameterization is given, the one forms $D_{6,7}$ vanish identically when $k_2=k_3$, reducing the dimension of the states space. Notice that this kind of degeneracy does not give rise to diagonal components of the symmetric logarithmic one form, because the determinant \eqref{U3degmatrix} does not vanish.

Therefore the form of $\dl\rho$ in this case is totally similar to the one presented for pure states (compare with \eqref{dlpure3}). This fact reflects the structure of the states space for this kind of degeneracies, that is to say the coset 
\begin{equation}
\qsp{U(3)}{U(2)\times U(1)}\simeq \mathbbm{CP}^2
\end{equation}

\subsection{Transversal direction}\label{transv}
Notice that in principle $d\rho$ can be defined on the whole Lie algebra, rather than on the orbit only, by allowing the weights $k_i$ to vary. In fact 
\begin{equation}
d^{tot}\rho=d^T\rho + d\rho
\end{equation}
where by $d^T\rho=\sum dk_i P_i$ we mean the differential along the { directions transversal to the orbit}. Then also the symmetric logarithmic one form has to change accordingly, and we have
\begin{equation}
d^{tot}\rho=d^T\rho + d\rho=\frac{1}{2}\{\dl^T\rho,\rho\} + \frac{1}{2}\{\dl\rho,\rho\}
\end{equation}
where we introduced the transversal logarithmic differential $\dl^T\rho$. We will see that for our purposes the two parts do not interact with each other, and the transversal direction (discussed for U(2) in \cite{es} and \cite{luati}) can be treated separately. { This fact originates from the existence of the normal bundle for the embedded submanifold $\iota(\P^{(m)}_n)\subseteq\mathfrak{u}(n)$. The \emph{$U$-transported} transversal derivative $\sum\partial_\theta k_i\rho_i$ is then some section in the normal bundle. A clarifying example \cite{es} is that of q-bits, where the space of states with fixed parameters is a sphere of radius $(2k_1-1)$, and the transversal direction is the radial degree of freedom $dk_1$}.

\section{Fisher Tensor} \label{se:ft}
The quantum Fisher information index is expressed in terms of the symmetric logarithmic derivative by the following formula:
\begin{equation}\label{fishind}
I(\theta)=\Tr{\rho(\dl\rho)^2}
\end{equation}
In what follows we will again drop the $\theta$ dependence and consider the whole orbit, which is a symplectic manifold, rather than its one dimensional real submanifold specified by $\rho= \rho(\theta)$.

Notice that for the discussion in \ref{transv} the logarithmic differential $\dl\rho$ splits into a transversal part and an horizontal part. It turns out that the Fisher information splits as well, as a matter of fact:
\begin{equation}\begin{aligned}
\Tr{\rho_0(\dl^{tot}\rho_0)^2}=&\Tr{\rho_0(\dl^T\rho_0)^2} + \Tr{\rho_0(\dl\rho_0)^2} + \\
+&\Tr{\rho_0\{\dl\rho_0,\dl^T\rho_0\}}
\end{aligned}\end{equation}
but last term vanishes because, from \eqref{structvanish}
\begin{equation}
\Tr{t_{\widehat{l}}\,\{t_{\bar{k}}\,,\,t_{\widehat{m}}\}}=0
\end{equation}

Following \cite{marmo,es}, we will not only consider the matrix form of $\dl\rho$, but rather promote it to a matrix valued one form, i.e. to a section of $\Omega^1(\mathcal{P}^{(m)}_n)\otimes\mathfrak{u}(n)$. This will allow us to define proper tensor field on the orbit $\mathcal{P}^{(m)}_n$, the \emph{Fisher Tensor}, as:
\begin{equation}\label{Fishertensor}
\mathfrak{F}=\Tr{\rho\dl\rho\otimes\dl\rho}
\end{equation} 
whose symmetric part has components equal to the above defined \eqref{fishind} Fisher information index.

It has been shown  \cite{marmo,chat,es} that this tensor agrees with the usual round metric on the sphere plus $i$ times the Berry phase connection both in the case of pure states in an arbitrary two-level system\footnote{This result can be generalized to n-dimensional pure states, \cite{brohug,marmo}.} and in the case of mixed states for a two-level system, when the space of quantum states is the manifold $\mathbbm{CP}^1\sim S^2$, Indeed, when the weights of the mixing are kept fixed, its symmetric and antisymmetric part are described in terms of the natural Kostant-Kirillov-Souriau symplectic form on co-adjoint orbits and yield a generalizations of the Fubini-Study metric for generic radii.

We would like to give a more precise geometrical interpretation of what has been found until today by looking at a different example, that of a rank-3 non-degenerate mixed state in a three-level system, whose space of states is no longer a projective manifold. This example will give us some interesting insights for a subsequent generalization of the results.

\subsection{$U(3)$ flag manifold}
As in the previous section, we start from the base point:
\begin{equation}
\rho_0=\mathsf{diag}(k_1,k_2,k_3)
\end{equation}
where the weights $k_i$ are all different, sum to one and are kept fixed.  We can reach any other maximum rank density matrix with the same eigenvalues via the $U(3)$ co-adjoint action: $\rho= U^\dagger \rho_0 U$. Clearly $\rho_0$ is kept fixed by the (diagonal) action of $U(1)^3$, so that we may identify the manifold of states with the coset space
\begin{equation}
\mathcal{P}^{(3)}_3=\qsp{U(3)}{U(1)^3}\simeq\{\rho\in\mathfrak{u}(3)\ \big|\ \rho=\Ad_U\rho_0,\ U\in U(3)\}
\end{equation}
which is a flag manifold inside the (dual) Lie algebra of the unitary group. It is a general fact that co-adjoint orbits of this kind are K\"ahler manifolds \cite{kir,berhol,pick,kir,borde}, the symplectic structure being the Kostant Kirillov Souriau equivariant form $\Omega_{KKS}$ \cite{kir}.

From its very definition, it is evident that the Fisher tensor is equivariant with respect to the $U(3)$ action, so that it is sufficient to compute it on the base point since the general definition on other points is done by simply acting with the Lie group adjoint representation. Expanding then $\rho_0$ as in \eqref{rhozeroexpansion} and $\dl\rho_0$ as in \eqref{basepointexpansion}, taking also into account eq. \eqref{equazioniU3}, after some algebra we find:
\begin{equation}
\mathfrak{F}=\Tr{\rho_0\left(L_i\lambda_i\right)\otimes \left(L_j\lambda_j\right)}
\end{equation}
where summation over repeated indices is understood. If we now notice that the following identity holds:
\begin{equation}\begin{aligned}
L_i\otimes L_j\lambda_i\lambda_j&=\sum_jL_j\odot L_j\left(\frac{2}{3}\id+ \sum_l f_{ijl}\lambda_l\right)\\
&+2\sum_{i<j}L_i\odot L_jf_{ijk}\lambda_k \\
&- 2\sum_{i<j}L_i\wedge L_jc_{ijk}\lambda_k
\end{aligned}\end{equation}
where $\odot,\wedge$ represent respectively the symmetrized and antisimmetrized tensor product, after some lengthy computations, one obtains:
\begin{equation}\begin{aligned}\label{FisherU3}
&\mathfrak{F}_{(3,3)}&=(k_1+k_2)(L_1^{\odot2} + L_2^{\odot2}) -2(k_1-k_2)L_1\wedge L_2\\
&&+ (k_1+k_3)(L_4^{\odot2} + L_5^{\odot2}) -2 (k_1-k_3)L_4\wedge L_5\\
&& + (k_2+k_3)(L_6^{\odot2} + L_7^{\odot2}) -2 (k_2-k_3)L_6\wedge L_7\\
\end{aligned}\end{equation}
It is interesting to observe that a similar calculation for a two-level system (see also an alternative derivation in \cite{es}) yields:
\begin{equation}
\mathfrak{F}_{(2,2)}=(k_1+k_2)(L_1^{\odot2}+L_2^{\odot2}) - 2(k_1-k_2)L_1\wedge L_2
\end{equation}
which, compared with (\ref{FisherU3}), again shows the underlying geometrical structure of $U(3)$ splitting in three exact copies of $SU(2)$.

We can see this more explicitly, by chosing a particular parametrization for $U(3)$. Consider now a group element $U$ that does not stabilize $\rho_0$. By direct exponentiation it may be found to be generated by the non-diagonal elements of the Lie algebra $\lambda_{\bar{k}}$:
\begin{equation}
U(z_1,z_2,z_3)=\mathrm{exp}\left\{i\sum_{\bar{k}}x_{\bar{k}}\lambda_{\bar{k}}\right\}
\end{equation}
with $z_1=x_1+ix_2, z_2=x_4+ix_5, z_3=x_6+ix_7$.  Infinitesimally we have
\begin{equation}
\id + i\sum_{\bar{k}}\epsilon_{\bar{k}}\lambda_{\bar{k}}=\id+i\widetilde{K}_0
\end{equation}
where we have set
\begin{equation}
i\widetilde{K}_0=\left(\begin{array}{ccc}
0 & \epsilon_2+i\epsilon_1 & \epsilon_5 +i \epsilon_4 \\
\epsilon_2 - i\epsilon_1 & 0 & \epsilon_7+i\epsilon_6 \\
\epsilon_5 -i \epsilon_4 & \epsilon_7 - i \epsilon_6 & 0
\end{array}\right)
\end{equation}
Then, the matrix component $\widetilde{d\rho}_0$ of the differential form $d\rho_0=-i[K_0,\rho_0]$ is found to be

\begin{equation}
\widetilde{d\rho}_0=\left(\begin{array}{ccc}
0 & r_1(\epsilon_2+i\epsilon_1) & r_2(\epsilon_5 +i \epsilon_4) \\
r_1(\epsilon_2 - i\epsilon_1) & 0 & r_3(\epsilon_7+i\epsilon_6) \\
r_2(\epsilon_5 -i \epsilon_4) & r_3(\epsilon_7 - i \epsilon_6) & 0
\end{array}\right)
\end{equation}
Where $r_1=k_1-k_2$, $r_2=k_1-k_3$ and $r_3=k_2-k_3$.

Promoting matrices to differential forms via 
\begin{equation}\epsilon_{2,5,7}+i\epsilon_{1,4,6}\mapsto dz_{1,2,3}^*
\end{equation}
we obtain
\begin{equation}
d\rho_0=\left(\begin{array}{ccc}
0 & r_1dz_1^* & r_2dz_2^* \\
r_1dz_1 & 0 & r_3dz_3^* \\
r_2dz_2 & r_3dz_3 & 0
\end{array}\right)
\end{equation} 
which has the expansion on the Lie algebra generators
\begin{equation}\label{drhoU3}\begin{aligned}
D_{1,4,6}=r_{1,2,3}\left(d\mathrm{Re}z_{1,2,3}-d\mathrm{Im}z_{1,2,3}\right)\\
D_{2,5,7}=r_{1,2,3}\left(d\mathrm{Im}z_{1,2,3}+d\mathrm{Re}z_{1,2,3}\right)\\
\end{aligned}\end{equation}
Formally, this promotion to differential forms is justified by the map $\ad_{\rho_0}\colon \mathfrak{u}(n)\longmapsto T\mathfrak{O}_{\rho_0}$ sending\footnote{Notice that the map is a bijection on the off diagonal part.} the parametrized Lie algebra $x_i\lambda^i$ into the curve in the tangent space 
\begin{equation}
X(x_i)=2x_i \rho_j c^{ijk}\lambda_k
\end{equation}
Here the \emph{differential}, or tangent, map promotes the parameters $x_i$ to differentials $dx_i$ so that the same element reads
\begin{equation}
d\rho\big|_0=2\rho_j dx_i\otimes\lambda_k
\end{equation}
where we emphasized the nature of one form with values in the Lie algebra of $d\rho$. 

Now we are ready to combine \eqref{equazioniU3} with \eqref{FisherU3} and the \eqref{drhoU3}'s to obtain the explicit Fisher tensor for a generic three-level threefold mixed quantum state, namely
\begin{equation}\begin{aligned}\label{Fish3,3}
\mathfrak{F}_{(3,3)}=4\frac{(k_1-k_2)^2}{(k_1+k_2)^2}&\{(k_1+k_2)dz_1\odot dz_1^*\\
&-i(k_1-k_2)dz_1\wedge dz_1^*\}\\
+4\frac{(k_1-k_3)^2}{(k_1+k_3)^2}&\{(k_1+k_3)dz_2\odot dz_2^*\\
&-i(k_1-k_3)dz_2\wedge dz_2^*\}\\
+4\frac{(k_2-k_3)^2}{(k_2+k_3)^2}&\{(k_2+k_3)dz_3\odot dz_3^*\\
&-i(k_2-k_3)dz_3\wedge dz_3^*\}\\
\end{aligned}\end{equation}
which is an even more explicit evidence of the splitting in three $SU(2)$ copies. Compare, for instance, with \cite{es} where the explicit expression for the Fisher tensor reads
\begin{equation}\begin{aligned}\nonumber
\mathfrak{F}_{(2,2)}&=4|\mu|^2\frac{(k_1-k_2)^2}{(k_1+k_2)^2}\{(k_1+k_2)dz\odot dz^*\\\label{Fish2,2}
&-i(k_1-k_2)dz\wedge dz^*\}
\end{aligned}\end{equation}
where $|\mu|^2$ is computed to be $|\mu|^2=(1+|z|^2)^{-2}$ due to the particular choice of coordinates.

\subsection{Reduction to degenerate orbits}
Given the expression on the whole flag manifold $\qsp{U(3)}{T^3}$ it is easy to recover the Fisher tensor on the degenerate submanifold of pure states $\qsp{U(3)}{U(2)\times U(1)}\simeq\mathbbm{CP}^2$ and in the subcases discussed in section \ref{3mixed}. As a matter of fact it is sufficient to impose the projection conditions $k_1=1,\ k_2=k_3=0$ or $k_3=0$ respectively to obtain
\begin{equation}
\mathfrak{F}_{\mathbbm{CP}^2}=4 dz_1^*\otimes dz_1 + 4 dz_2^*\otimes dz_2
\end{equation}
in the pure, projective case while in the intermediate situation one has
\begin{equation}\begin{aligned}
\mathfrak{F}_{k_3=0}=4\frac{(k_1-k_2)^2}{(k_1+k_2)^2}&\{(k_1+k_2)dz_1\odot dz_1^*\\
&-i(k_1-k_2)dz_1\wedge dz_1^*\}\\
+4k_1dz_2\otimes &dz_2^* + 4k_2dz_3\otimes dz_3^*
\end{aligned}
\end{equation}
Notice that this last case is not formally different from the generic case with the condition that $k_i\not=k_j$ for $i\not=j$. The dimension of the states space  is still 6 (real) but the vanishing of $k_3$ makes the definition of $\dl\rho$ dependent on a choice.

Notice, moreover, that if one were to consider the solely remaining degeneracy possibility, that is to say the combination $k_2=k_3\not=0$, he would end up with a configuration totally similar to the pure state case, but with a different weighting. As a matter of fact, enforcing this condition on \eqref{Fish3,3}, one remains with
\begin{equation}\begin{aligned}
&\mathfrak{F}_{k_2=k_3}=4\frac{(k_1-k_2)^2}{(k_1+k_2)^2}\,\times\\
&\{(k_1+k_2)dz_1\odot dz_1^* -i(k_1-k_2)dz_1\wedge dz_1^*\\
&\phantom{\{}(k_1+k_2)dz_2\odot dz_2^* -i(k_1-k_2)dz_2\wedge dz_2^*\}
\end{aligned}\end{equation} 

All of this formulae can be recovered also by computing the correct symmetric logarithmic form and then using formula \eqref{Fishertensor}. It is just a matter of lengthy but straightforward calculations to show that the two procedures yield to the same results when the degeneracy equations are trivially solved (that is to say when $\dl\rho$ is off diagonal). In other words the trivial solution $L_{\widehat{i}}=0$, with $\widehat{t}_{\widehat{i}}$ a diagonal generator, is compatible with the explicit degeneration procedure that we applied in the current section. 

The Fisher tensors $\mathfrak{F}(k_1,k_2,k_2)$ are therefore elements of a family depending on the mixing parameters, such that we may recover the degenerate Fisher tensors either by computing the proper degenerate symmetric logarithmic one form in the trivial non-diagonal choice, or by sending the mixing parameters to their degenerate values explicitly.

\section{Conclusions and further developments}
With the computations performed in section \ref{formula} we provided a way to obtain the symmetric logarithmic derivative any time that the structure constants of the unitary group under consideration are known. This reduces the problem to the computation of those constants for general dimensions, a problem that can be attacked possibly via automatic computations, once the construction of generalized Pauli matrices is taken into account (see for instance \cite{sarid} and \cite{Maki}). { The present method is sensibly different from the original (formal) solution proposed in \cite{brau}, in that it doesn't require the decomposition of the derivative $\partial\rho$ onto the eigenstates $\psi_i$. As a matter of fact we avoid finding the explicit eigendecompositions of expressions like $\partial\psi_i=a\psi_i + b\psi_i^\bot$. 

The expansion in the Lie algebra generators, instead, allows to obtain an explicit expression for $d\rho$ given a parameterized expression for $\rho$. Most importantly, it allows to simply solve the problem above the reference point $\rho_0$, and obtain the expressions above all other points $\rho$ by conjugation.

The price to pay, the \emph{computational} complexity stands in the computation of the structure constants for $U(n)$ and a parameterization of the orbits $\mathcal{P}^{(m)}_n$. We stress, once more, that this problem is of general interest and might be automated through the Bruhat coordinatization procedure (essentially a clever Gram-Schmidt construction) presented, for instance, in \cite{pick}. }

The { resulting} uniqueness of the definition of such an object {in the case of} flag manifolds is surely suggesting that the role of the symmetric logarithmic one form is more profound than what originally imagined. Therefore, the study of its geometric nature and that of the Fisher tensor, acquires a strong charm both for the physicist and for the mathematician.

It is reasonable to think that due to the property of being uniquely defined, the one form $\dl\rho$ plays an important role in the fiber bundle of states, dictating how this definition must be extended in the degenerate orbits. Moreover, being it the key ingredient to construct the Fisher tensor, one could expect the latter to have a clear intrinsic interpretation as well as important geometric properties, as it happens for q-bits.

\section{Acknowledgements}
M.S. would like to thank Ivan Contreras and Camilo Arias Abad for interesting and helpful discussions. 

M.S. also acknowledges partial support of SNF Grant No. PDFMP2\_137103/1.


\vspace{20pt}


































\begin{thebibliography}{99}

\bibitem{erc} E. Ercolessi, G. Marmo and G. Morandi, La Rivista del Nuovo Cimento {\bf 33}, 401 (2010) 
\bibitem{berry} M.V. Berry, Proc. Roy. Soc. {\bf A 392}, 45 (1984) 
\bibitem{aa} Y. Aharonov and J. Anandan,  Phys. Rev. Lett. {\bf 58}, 1593 (1987) 
\bibitem{rez} A.T. Rezakhani and P. Zanardi, Phys. Rev. A {\bf 73}, 012107 (2006)  
\bibitem{bro} D.C. Brody and L.P. Hughston, J.Geom.Phys. {\bf 38}, 19 (2001) 
\bibitem{beng} I. Bengtsson and K.  \.Zyczkowski, {\it Geometry of quantum states: an introduction to quantum entanglement}. Cambridge University Press (2006)
\bibitem{brau} S.L. Braunstein and C.M. Caves, Phys. Rev. Lett. {\bf 72}, 3439 (1994) 
\bibitem{brauII} S.L. Braunstein, C.M. Caves and G.J. Milburn, Ann. Phys. {\bf 247}, 135 (1996)
\bibitem{hel} C.W. Helstrom, Phys. Lett. A {\bf 25}, 101 (1967) 
\bibitem{hol} A.S. Holevo 1982 {\it Probabilistic and Statistical Aspects of Quantum Theory} (Amsterdam: North-Holland) Russian original, 1980
\bibitem{barn} O.E. Barndorff-Nielsen and R.D. Gill, J. Phys. A {\bf 33}, 4481 (2000) 
\bibitem{luati} A. Luati, Indian J. Stat. {\bf 70} (2008) 25  and The Ann. Stat.  {\bf 32}, 1770 (2004)
\bibitem{es} E. Ercolessi and M. Schiavina, J. Phys. A: Math. Theor. {\bf 45}, 365303 (2012)
\bibitem{rao} C.R. Rao, Bull.Calcutta Math.Soc. {\bf 37}, 81 (1945) 
\bibitem{bur} J. Burbea and C.R. Rao, Probab. Math. Stat. {\bf 32}, 41 (1984) 
\bibitem{bro2} D.C. Brody, J.Phys.A:Math.Theor. {\bf 44}, 252002 (2011) 
\bibitem{bro3} D.C. Brody and L.P. Hughston, Proc. R. Soc. A {\bf 454} (1998) 2445 and Proc.R.Soc. A {\bf 455} (1999) 1683-17
\bibitem{brohug} Brody D.C. and Hughston L.P. J. Geom. Phys. {\bf 38}, 19 (2001)
\bibitem{petz} Petz, D. Linear Algebra Appl. {\bf 244}, 81 (1996) 
\bibitem{zyc} K. \.Zyczkowski and W. Slomczy\'nski, J. of Phys. A: Math. Gen. {\bf 34}, 6689 (2001) 
\bibitem{chat} S. Chaturvedi, E. Ercolessi, G. Marmo, G. Morandi, N. Mukunda and R. Simon, European Physical Journal C {\bf 35}, 413 (2004) 
\bibitem{marmo} P. Facchi, R. Kulkarni, V.I. Man'ko, G. Marmo, E.C.G. Sudarshan and F. Ventriglia, Physics Letter A {\bf 374} (2010) 4801
\bibitem{Fuji} A. Fujiwara and H. Imai, J. Phys. A: Math. Theor. {\bf 41}, 255304 (2008)
\bibitem{pick} R.F. Picken, J. Math. Phys. {\bf 31}, 616 (1990)
\bibitem{sarid} U. Sarid, J. Math. Phys. {\bf 26}, 1921 (1985)
\bibitem{Maki} Z. Maki, T. Maskawa, and I. Umemura, {\it Quartet scheme of hadrons in 
chiral U(4)XU(4)}, Prog. Theor. Phys. {\bf 47}, 1682 (1972)
\bibitem{kir} A.A. Kirillov, \emph{Lectures on the orbit method}, volume 64 of \emph{Graduate studies in mathematics}, American Mathematical Society, 2004
\bibitem{emm} E. Ercolessi, G. Marmo and G. Morandi, International Journal of Modern Physics A, {\bf 16}, 31 (2001) 5007-5032
\bibitem{berhol} J. Bernatska and P. Holod, {\it Geometry and Topology of coadjoint orbits of semisimple Lie groups} in Ninth International Conference on Geometry, Integrability and Quantization (2007) 1-21
\bibitem{borde} M. Bordemann, M. Forger and H. R\"omer, Commun. Math. Phys. {\bf 102}, (1986) 605-647
\end{thebibliography}
\end{document}